\documentstyle[prb,preprint,aps]{revtex}

\begin{document}                
\draft

\title{Realistic Electron-Electron Interaction
in a Quantum Wire}

\author{Krzysztof Byczuk}
\address{Institute of Theoretical Physics, Warsaw University,
Ho\.za 69, PL-00-681 Warsaw, Poland; and \\ Lyman Laboratory of
Physics, Harvard University, MA 02138, USA}

\author{Tomasz Dietl}
\address{Institute of Physics, Polish Academy of Sciences, 
al. Lotnik\'ow 32/46, PL-02-668 Warszawa, Poland}

\date{\today}
\maketitle
\begin{abstract}
The form of an effective electron-electron interaction in a
quantum wire with a large static dielectric constant is
determined and the resulting properties of the electron liquid
in such a one-dimensional system are described.  The exchange
and correlation energies are evaluated and a possibility of a
paramagnetic-ferromagnetic phase transition in the ground state
of such a system is discussed.  Low-energy excitations are
briefly described.  
\end{abstract}


In  three and  two dimensional systems any interaction between
electrons (in a forward scattering channel) appears to have a
marginal property.\cite{1} This means that the low-energy
dynamics of such  systems is described by the Fermi liquid
theory where quasi-particles are in a one-to-one correspondence
with bare electrons.\cite{landau} One-dimensional systems are
very exceptional in this respect.  The interaction between
fermions is relevant and leads to a new collective type of  a
particle motion at low-energies.  Such a system is described by
the Luttinger liquid theory, according to which  a spin-charge
separation and an anomalous scaling take place.\cite{2}

A recent progress in nanofabrication enables to construct
various one-dimensional structures, in which the Luttinger
liquid theory can be experimentally tested.\cite{general} In
such structures the effective interaction between the electrons
is not a short-range one  as in the standard Tomonaga-Luttinger
(TL) model.\cite{2} Rather it must be of a long-range type
because it originates from the Coulomb electrostatic forces
between the charged particles.  Accordingly, in such a case
certain properties of the Luttinger liquid will be modified.

In this Brief Report we analyze theoretically the form of an
effective interaction between the electrons in a one-mode
quantum wire, and the corresponding new properties of the
system.  In particular, we are interested in how ground-state
and excitation energies will be affected by a difference between
the dielectric constants of the wire and its environment. Such a
difference is large in the case of free standing wires,
particularly of IV-VI semiconductors such as PbTe, characterized
by a large static dielectric constant $\epsilon \approx
1000$.\cite{cos} This very large dielectric constant is due to
the proximity of this system to the ferroelectric phase
transition.  In those cases the electrostatic potential between
the electrons is strongly modified because of the presence of
image charges that assure the correct boundary conditions.

In order to determine the form of the interaction in such a
system we consider a model, in which the electrons can propagate
along an infinitely long cylinder of radius $a$, made of a
material with a macroscopic dielectric constant $\epsilon_1$.
This cylinder is embedded into a bulk system with a macroscopic
dielectric constant $\epsilon_2$.  Further, we assume that the
single-electron wave function vanishes at the boundary of this
cylinder.  In other words, an electron moves in a cylindrical
potential well with infinite barriers.  Eigenfunctions
$\Psi_{nmk}$ in such a geometry are readily to find, and the
result in the cylindrical coordinates $(\rho, \phi, z)$ is,
\begin{equation}
\Psi_{nmk}(\rho,\phi,z) = \frac{1}{a\sqrt{\pi} J_{m+1}(X_{mn})}
J_m\left(\frac{X_{mn}}{a}\rho\right) e^{im\phi} \frac{e^{\pm
iqz}}{\sqrt{L}},
\label{1}
\end{equation}
where $J_m(x)$ is the Bessel function,\cite{smyth} $X_{mn}$ are
nodes of $J_m(x)$, $(n,m,q)$ is a set of quantum numbers.

In the following we are interested in properties of the
quantum wire when only the lowest quantum level ($n=m=0$) is
occupied.  In this case the radial dependence of the wave
function (1) can be approximated very well by the following
parabola \cite{gold}
\begin{equation}
\psi(\rho) = \left\{
\begin{array}{ccc}
\sqrt{\frac{3}{\pi a^2}} (1- (\frac{\rho}{a})^2) & \mbox{for} &
\rho < a \\
0 & & \mbox{otherwise.}
\end{array}
\right.
\label{2}
\end{equation}

An electrostatic potential $V({\bf r} - {\bf r}_0)$ between two
electrons is determined by solving a Poisson equation $\nabla^2
V = \frac{-4 \pi e}{ \epsilon_1}
\delta({\bf r}-{\bf r}_0)$, where $e$ is an charge of an electron,
with  appropriate boundary conditions at $\rho=a$, i.e.,
$V_1=V_2$ and $\epsilon_1 \partial_{\rho} V_1 =
\epsilon_2\partial_{\rho} V_2$.\cite{smyth} As a result we find
that
\begin{eqnarray}
V({\bf r} - {\bf r}_0) =
\frac{e}{ \epsilon_1}
\frac{2}{\pi}
\sum_{m=-\infty}^{\infty} \int_0^{\infty} d \lambda
e^{im (\phi -\phi_0) } \cos[\lambda(z-z_0)] \left[\frac{}{}
I_m(\lambda \rho_<) K_m(\lambda \rho_>) + \right.  \nonumber \\
\left.
\frac{
( 1- \frac{ \epsilon_2 }{ \epsilon_1 } ) K_m'(\lambda a)
K_m(\lambda a) }{
\frac{ \epsilon_2 }{ \epsilon_1 }
K_m'(\lambda a) I_m(\lambda a) - K_m(\lambda a) I_m'(\lambda a)
} I_m(\lambda \rho)
\right],
\end{eqnarray}
where $\rho_{<} \equiv \min(\rho,\rho_0)$,  $\rho_{>} \equiv
\max(\rho,\rho_0)$,  $I_m(x)$, $K_m(x)$ are the modified
Bessel functions,\cite{smyth} and primes denote their
derivatives.  We have not considered here effects coming from
 electrodes, which are usually attached to the system,
assuming that they are very far from each other and  modify the
system properties only very close to the edges.  However, for
short wires and quantum point contacts the electrodes may be
important as well.

An effective Hamiltonian describing  electrons in a 1D quantum
wire with the lowest level occupied only ($n=m=0$) 
has a two-body matrix element 
(determined in a one-particle basis of states (2)) in the
 following form:
\begin{eqnarray}
V(aq) = \frac{36}{\pi}\frac{e^2}{\epsilon_1}\frac{1}{|aq|}\left[
\frac{1}{10} - \frac{2}{3}\frac{1}{|aq|^2} +
\frac{32}{3}\frac{1}{|aq|^4}-
\frac{64}{|aq|^4} I_3(|aq|) K_3(|aq|) \right] + \label{6} \\
\frac{6}{\pi} \frac{e^2}{\epsilon_1} \frac{A_0(|aq|)}{|aq|}
\left[ I_1(|aq|) - \frac{4}{|aq|}I_2(|aq|) +
\frac{8}{|aq|^2}I_3(|aq|) - 2 I_3(|aq|) +
\frac{8}{|aq|}I_4(|aq|) + I_5(|aq|) \right],
\nonumber
\end{eqnarray}
where
\begin{equation}
A_0(|aq|)=
\frac{
( 1- \frac{ \epsilon_2 }{ \epsilon_1 } ) K_0'(|aq|) K_0(|aq|)
}{
\frac{ \epsilon_2 }{ \epsilon_1 }
K_0'(|aq|) I_0(|aq|) - K_0(|aq|) I_0'(|aq|) }.
\end{equation}
The first part, which was found previously in Ref.~7,
corresponds to the long range Coulomb interaction between the
electrons moving in a quasi-1D constriction.  The second part,
which disapears when $\epsilon_1=\epsilon_2$, describes the
interaction between an electron and the image charges, which
assures the proper  boundary conditions.

In Fig.~1, we plotted the matrix element $V(aq)$ as a function
of $aq$ for different ratios $\epsilon_1/\epsilon_2=1,10,100$.
We see that $V(aq)$ is a decreasing function of $aq$ and for
large values of $aq$ becomes relatively small.  Additionally, we
find that the numerical values of $V(aq)$ diminish when the
dielectric constant $\epsilon_1$ increases and this reduction is
very different when the image charges are taken into account as
is shown in the inset to Fig.~1.  For small $aq$, however, the
effective interaction is seen to diverge.  We have been able to
examine this limit analytically and found that at $|aq|
\ll 0$ the matrix element behaves as
\begin{eqnarray}
V(aq)= \frac{e^2}{\pi} \frac{1}{\epsilon_2} \left\{ -\ln \left|
\frac{aq}{2}\right|
 + \left[ \gamma \left(2 \frac{\epsilon_2}{\epsilon_1}-1\right)
-
\frac{73}{120} \frac{\epsilon_2}{\epsilon_1} \right] + \nonumber
\right.  (aq)^2 \left[
\frac{\gamma^2}{2}\left(1-\frac{\epsilon_1}{\epsilon_2}
\right) - 
\frac{\gamma}{16}\left(1+ \frac{\epsilon_2}{\epsilon_1}\right) +
\frac{1}{4}- \right.\\
\left.
\frac{89}{840}\frac{\epsilon_2}{\epsilon_1}+
\left.
\left(
\gamma \left(1-\frac{\epsilon_1}{\epsilon_2}\right)
 - \frac{1}{16} \left( 1+ \frac{\epsilon_2}{\epsilon_1}\right)
\right) \ln \left|\frac{aq}{2}\right|+
\frac{1}{2} \left(1-\frac{\epsilon_1}{\epsilon_2}\right) 
\ln^2 \left| \frac{aq}{2}\right| \right]
 \right\} ;
\label{8}
\end{eqnarray}
so it diverges logarithmically ($\gamma$ is the Euler constant).
This behavior is characteristic for a 1D Fourier transform of
the Coulomb interaction $e^2/r$ (with $a$ as a short-distance
cut off).\cite{gold,shultz} However, it is surprising that
$V(aq)$ for $ |aq|
\rightarrow 0$ {\em does not depend} on the dielectric constant
$\epsilon_1$ of the wire but only on the dielectric constant
$\epsilon_2$ of the environment.  This means that as the wire is
infinitely thin (or particles are very far from each other) a
single electron interacts mainly with the image charges and not
directly with the other electrons in the wire.  We have checked
explicitly that the disapearence of the dielectric constant
$\epsilon_1$  in the first leading term takes place for both
wave functions (\ref{1}) and (\ref{2}), which are separable in
the cylindrical coordinates.\cite{note}

Having determined the analytical form of the electron-electron
interaction matrix element we can calculate ground state
properties of such a quantum wire.  We assume that the electrons
propagate in a jellum environment with positive ion charges.
This leads to the cancelation of a direct self-energy
contribution, which otherwise brings about infinities in the
perturbation expansion.\cite{fetter} Hence, the Hartree-Fock
correction to the ground state energy has the simple form
\begin{equation}
E_{HF}^x=-2\int_{-k_F}^{k_F} \frac{dk}{2 \pi}
\int_{-k_F}^{k_F} \frac{dp}{2 \pi} V(a|k-p|),
\end{equation}
and its numerical values are depicted in Figs.~2a and 2b, for
the different radii ($a$) of the wire and the different RPA
parameters $r_s\equiv(2 a_0^{\ast} n)^{-1}$, where $n$ is the
density of the 1D electron gas.\cite{uwaga} All results below
are presented in the atomic units where $a_0^{\ast} =
\frac{\epsilon_2}{m^{\ast} e^2}$ is the effective Bohr radius,
and $Ry^{\ast} = \frac{m^{\ast} e^4}{2 \epsilon_2^2 }$ is the
effective energy unit; so-called the effective Rydberg.  Note
that since in the leading term of the expansion (\ref{8}) only
the dielectric constant $\epsilon_2$ appears, we have defined
the effective Rydberg and the effective Bohr radius with
$\epsilon_2$.  We see from Figs.~2 that, as expected, the
absolute value of the exchange energy is smaller for either a
wider wire ($a=10$), a less dense electron gas in the wire
($r_s=4$), and a  greater dielectric constant $\epsilon_1$.
However, the decay of $E_{HF}^x$ with $\epsilon_1$ is
significantly weaker in our model (\ref{6}) than in the model
of Ref.~7, in which no image charges were considered,
$A_0=0$, as is shown in the inset to Fig.~2a.  This comaprision
convincingly demonstrates the important influence of the image
charges upon the ground state energy of the wire.

We have also evaluated a correlation energy $E_c$, {\it i.e.},
the correction to the Hartree-Fock energy originating from a
linear screening of the electron-electron interaction due to
electron-hole excitations.  For that we calculate the effective
(RPA) interaction, which in the static limit has the
form\cite{fetter} $ V^{eff}_{RPA}(aq) = V(aq)/\epsilon(aq), $
with the dielectric function $\epsilon(aq)$ approximated as $
\epsilon(aq) \approx 1 + 2 V(aq) /\pi v_F,
$ where $v_F=\frac{\hbar k_F}{m^{\ast}}$ is the Fermi velocity.
The correlation energies $E_c$ are plotted in Figs.~3a and 3b
for the various wire's radii $a$ and the parameters $r_s$.

In order to estimate the transition line between the
paramagnetic and the ferromagnetic ground states we compare the
ground state energies for these two phases.\cite{gold96} Our
results are shown in Fig.~4 for the two  wire's radii $a=1$ and
$10$.  The critical value of the RPA parameter $r_s^c$, above
which the system would be completely polarized, increases with
increasing $\epsilon_1$. This increase is, however, sublinear
which indicates that the effects of the Coulomb carrier-carrier
interaction may remain important even in nanostructures with a
large bulk dielectric constant.  At the same time, if there was
$\epsilon_1 \ll \epsilon_2$ then the critical value of $r_s^c$
would go rapidly to zero (c.~f. the inset to Fig.~4).  This result
suggests that if a quantum wire with a small dielectric constant
was deposited on a ferroelectric substrate, then the
ferromagnetic phase instability would be even more likely.

It is tempting at this point to make a comment on the so-called
$0.7$ step in the quantized conductance, which is observed in
the quantum point contacts of GaAs/AlGaAs,\cite{thomas98} and
also in PbTe.\cite{cos} Our results make possible to evaluate
electron concentrations at which the zero-temperature
ferromagnetic instability might appear. However, we cannot
exclude other types of instabilities, e.g., the charge- or the
spin-densities-waves, which are not discussed in the present
paper.

Finally, we  briefly describe the low energy excitations in our
model.  As is known, arbitrary weak interaction destroys the
Fermi liquid description in one dimensional systems.  Instead,
the Luttinger liquid theory emerges as the proper low-energy
principle in this case.\cite{2}

The one-dimensional model of interacting electrons can be solved
exactly in the low-energy limit by means of a bosonization
\cite{2}.  In our case we must linearized the dispersion
relation $\epsilon_k-\epsilon_F= v_F(\pm k-k_F)$, where $v_F=
\hbar k_F/m^{\ast}$ is the Fermi velocity at the two Fermi
points $\pm k_F$.  Next, we introduce the so-called left and
right moving operators corresponding to $\pm v_F$, respectively,
and then define fluctuation density operators
$\hat{\rho}_{q\sigma\alpha} = \sum_k
c_{k+q\sigma\alpha}^{\dagger}c_{k\sigma\alpha}$ for each branch
$\alpha=R,L$ separately.  In terms of these operators the many
body Hamiltonian is bilinear and can be diagonalized exactly
because in the low-energy limit $\hat{\rho}_{q\sigma\alpha}$
obeys boson-like commutation relation.\cite{2}

As a result we find that spectra in the charge and the spin
channels are different.  Namely, the eigenvalue in the charge
sector of the many-body theory is $\omega^c_q = v_F
q\sqrt{1+2V(aq)/\pi v_F}$, whereas the spin degrees of freedom
propagate with the free dispersion relation $\omega^s_q=v_Fq$.
We see that in the low-energy limit these degrees of freedom are
completely separated as in TL model.  However, in the present
case the charge excitation energy is not a linear function of
$q$ because the interaction behaves as $V(aq) \sim
\frac{1}{\epsilon_2}\ln|aq|$ for $|aq|\rightarrow 0$.
Nevertheless, $\omega^c_q \rightarrow 0$ as $q\rightarrow 0$.
The spin degrees of freedom are not affected by the interaction
since $V(aq)$ only couples the charge density fluctuations.

The model with the logarithmic divergence of $V(aq)$ leads to
non-analytic properties of thermal quantities.  For example, the
specific heat of this system at low temperatures is $C = \gamma
T + \beta T \ln T$ contrasting with the standard result in TL
model where $C_{TL}=\gamma_{TL} T$.\cite{2} Additionally, a
single-particle density of states vanishes at the Fermi level as
$N(\omega) \sim \omega^x \ln^y\omega$ with $x$ and $y$ being
non-universal constants.  Again in TL model $N_{TL}(\omega) \sim
\omega^{\mu}$.\cite{2} Also, as shown  by
Schultz,\cite{shultz} the long range correlation functions have
logarytmic corrections, and this might drive the system into a
Wigner crystal.

In conclusion, these results strongly suggest that the ground
state and the  low-energy properties of 1D electrons in a
quantum wire with the realistic form of the interaction are very
different from those expected in the framework of the standard
TL-type models.  In particular, the actual form of the potential
$V(aq)$, and particularly, the screening effects due to the
boundaries should modify transport properties in this system. It
would, therefore, be very interesting to evaluate directly the
conductivity.

It is a pleasure to acknowledge discussions with G.~Bauer and
G.~Grabecki.  This work was supported by the Committee for
Scientific Research (KBN) of Poland through Grant No
2-P03B-6411.  KB is also very grateful to B.~Halperin for his
hospitality at Harvard University where the final part of this
work was completed.  This visit is sponsored by 
the Foundation for Polish Science (FNP).


\begin{figure}
\caption{Fourier transform of the electrostatic interaction,
$V(aq)$ as a function of $aq$ for different ratios
$\epsilon_1/\epsilon_2$ of the dielectric constants inside and
outside the wire of the radius $a$. The inset shows $V(aq)$ as a
function of $\epsilon_2/\epsilon_1$ for $|aq|=0.1$ in our model
(\ref{6}) (solid line) and in the model without the image
charges (dashed line).}
\end{figure}

\begin{figure}
\caption{ Exchange energy as a function of
$\epsilon_1/\epsilon_2$ for different: a) radius $a= 1, 2, 5,
10$ with $r_s=1.0$, b) RPA parameters $r_s=0.5, 1, 2, 4$ (from
bottom to top) with $a=1.0$.  Energy and length units are given
in the units of the effective Rydberg and Bohr radius calculated
with the effective mass $m^*$ inside the wire and the dielectric
constant $\epsilon_2$ outside the wire. The inset compares our
model (Eq.~\ref{6}) (solid lines) with the model without the
image charges (dashed lines).}
\end{figure}

\begin{figure}
\caption{ Correlation energy as a function of
$\epsilon_1/\epsilon_2$ for different: a) radius $a=1, 2, 5, 10$
with $r_s=1.0$, b) RPA parameters $r_s=0.5, 1, 2, 4$ with
$a=1.0$.}
\end{figure}

\begin{figure}
\caption{
Critical value of the RPA parameter $r_s$ as a function of
$\epsilon_1/\epsilon_2$ for $a=1,10$.  Above $r_s^c$ the
ferromagnetic phase is a stable ground state.  Inset shows the
behavior of $r_s^c$ for small $\epsilon_1 \le 1$.}
\end{figure}


\begin{references}
\bibitem{note} We have also investigated the form of the
two-body matrix element in a general case when several modes are
occupied in the wire.  It turns out that as long as the particle
scattering between different modes $n$ ( with a change of
$m$-th quantum number, i.e. $z$-th components of the orbital
momentum) are forbidden, the form of the potential for small
$|aq|$ is $V(aq) = - \frac{e^2}{\pi \epsilon_2} \ln |aq| $; this
is effectively 1D potential.  Otherwise $V(aq)$ diverges with
different natural powers of $|aq|$.
\bibitem{uwaga} To accelerate the numerical calculations we have
approximated the first term in Eq.~(\ref{6}) by a function
$\frac{e^2}{\pi \epsilon_1}K_0(|aq|)$ that has a proper behavior
for small $|aq|$. Discrepancies for large $|aq|$ are less
significant since the integrals are cutoff at $\pm k_F$.
\end{references}
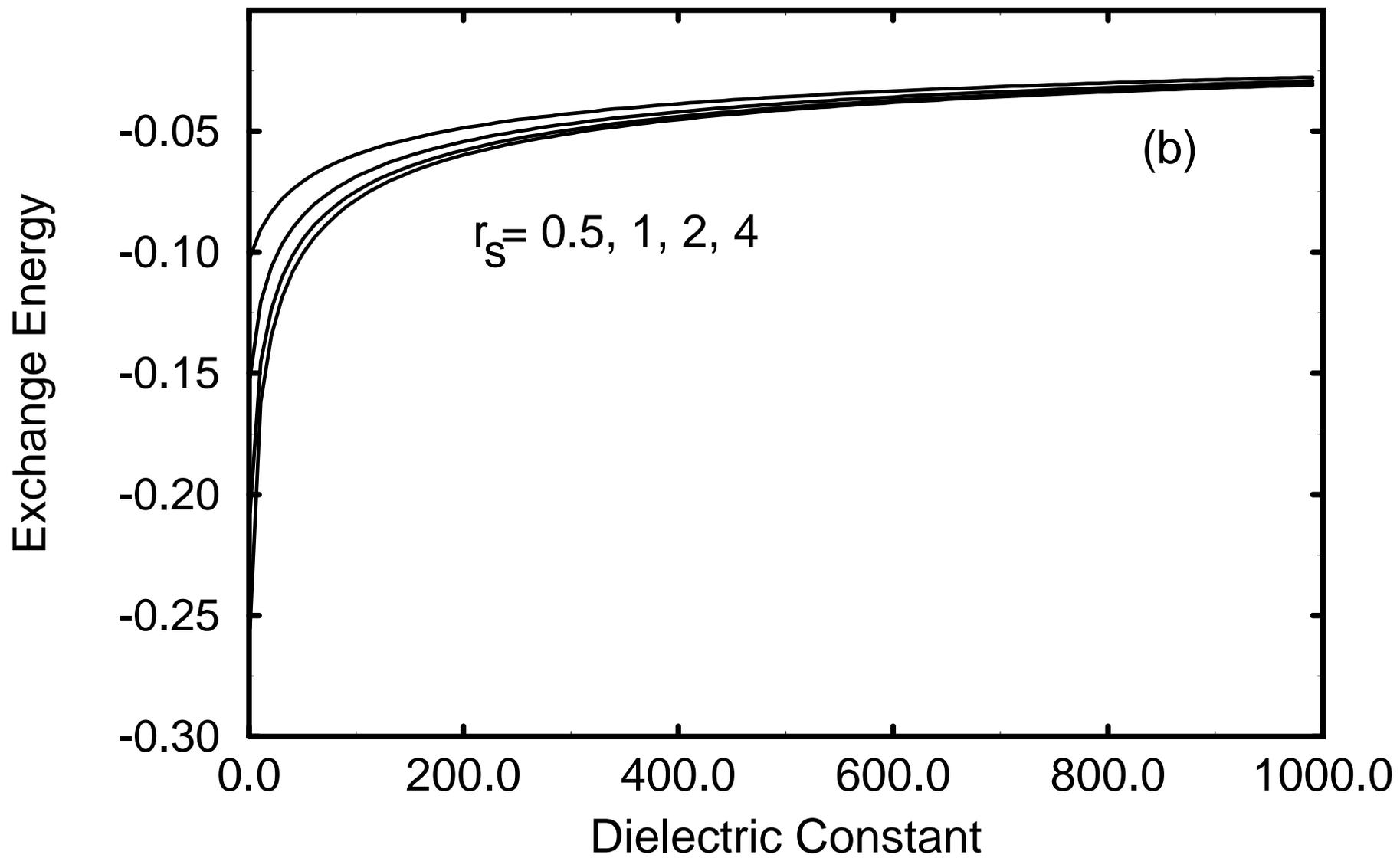
\end{document}